\documentclass[fleqn,usenatbib]{mnras}

\def\kms{\,km~s$^{-1}$}
\def\h2o{H$_2$O}
\def\nh3{NH$_3$}
\def\vlsr{$V_{\mbox{\scriptsize LSR}}$\ }

\def\g188{G188.95$+$0.89}
\def\co{$^{12}$CO}
\def\13co{$^{13}$CO}
\def\c18o{C$^{18}$O}
\def\ch{CH$_3$OH}

\usepackage[T1]{fontenc}

\usepackage{graphicx}	
\usepackage{amssymb}	
\usepackage{amsmath}	

\usepackage{subcaption}
\usepackage{float}
\pdfoutput=1
\usepackage{newtxtext,newtxmath}

\title[Massive Protocluster of a Periodic Maser Source G188.95$+$0.89]{Massive Protocluster of a Periodic Maser Source G188.95$+$0.89}

\author[M. M. Mutie et al.]{M.  M. Mutie$^{1,2}$,\thanks{E-mail: martmulesh@gmail.com (MMM)}
J. O. Chibueze$^{3,4}$\thanks{E-mail: james.chibueze@nwu.ac.za (JOC)},
K. El Bouchefry$^{5}$,
G. C. MacLeod$^{6,7}$,
\newauthor
J. Morgan$^{3}$
and P. Baki$^{1}$
\\
$^{1}$Department of Physics, Technical University of Kenya, P. O. Box 52428-00200, Nairobi, Kenya\\
$^{2}$Department of Physical Sciences, Chuka University, P. O. Box 109-60400, Chuka, Kenya\\
$^{3}$Centre for Space Research, Physics Department, North-West University, Potchefstroom 2520, South Africa\\
$^{4}$Department of Physics and Astronomy, University of Nigeria, Carver Building, 1 University Road, Nsukka, 410001, Nigeria\\
$^{5}$South African Radio Astronomy Observatory,Rosebank, Johannesburg, South Africa\\
$^{6}$The University of Western Ontario, 1151 Richmond Street. London, ON N6A 3K7, Canada\\
$^{7}$Hartebeesthoek Radio Astronomy Observatory, PO Box 443, Krugersdorp, 1741, South Africa
}

\date{Accepted 2021 June 30. Received 2021 June 30; in original form 2021 March 30}

\pubyear{2021}

\begin{document}
\label{firstpage}
\pagerange{\pageref{firstpage}--\pageref{lastpage}}
\maketitle

\begin{abstract}
We report the results of ongoing monitoring  of the 6.7\,GHz \ch~masers associated with \g188. In these observations five features are periodically varying and at least two exhibit evidence of velocity drifts. It is not clear the cause of these velocity drifts. The spectra have varied significantly since detection in 1991. The 11.45\kms\/  feature has decreased exponentially from 2003. Complementary ALMA 1.3\,mm continuum and line observational results are also presented. Eight continuum cores (MM1 -- MM8) were detected in \g188. We derived the masses of the detected cores. \g188~MM2 was resolved into 2 continuum cores (separated by 0.1\,\arcsec) in ALMA band 7 observations. Also CH$_3$OH  (4$_{(2,2)}$-3$_{(1,2)})$ thermal emission associated with MM2 is double peaked. We propose the presence of multiple (at least binary) young stellar objects in MM2. SiO emission exhibit a bow-shock morphology in MM2 while strong emission of $^{12}$CO at the east and west of MM2 suggest the presence of an east-west bipolar outflow.

\end{abstract}




    
    

\section{Introduction}
High mass star-forming regions (HMSFRs) is an ongoing topic of open debate as it relates to mass growth \citep{motte2018high}. The timescales of phases associated with this process they state are of the order of 10$^4$--10$^5$ years. Observation of such phases is very difficult. A unique property to HMSFRs is the presence of class II methanol (\ch) masers in their early formative phases \citep{ caswell1995variability,breen2013confirmation}. These class II \ch~masers have proven to be a reliable signpost of the very early stages of high-mass star formation \citep{menten1991discovery,caswell1996new}.
Recent results of monitoring related infra-red sources and class II methanol (\ch) masers towards HMSFRs have identified significant accretion events. At present only four accretion events have been detected, they are: S255IR-NIRS3 \citep{Fujisawa2015S255,Caratti17S255,Szymczak2018S255}, NGC6334I \citep{Hunter2017NGC,hunter2018extraordinary,macleod2018masing}, G323.46-0.08 \citep{proven2019discovery, MacLeod2021g323}, and G358.93-0.03 \citep{Sugiyama2019g358, MacLeod2019g358,Breen19G358,MacLeod2019g358,Burns20G358}.   

Another important phenomena associated with HMSFRs are periodic methanol masers discovered by  \citet{goedhart2003periodic}. Periodic masers are rare, there are only twenty-five known so far
\citep{goedhart2003periodic,goedhart2004long, goedhart2009short,goedhart2014,Araya2010period,szymczak2011periodic,Fujisawa2014periodic,maswanganye2015new,Szymczak2015period,Sugiyama2015periodic,maswanganye2016discovery,szymczak2016discovery,Sugiyama2017periodic,Szymczak2018period,proven2019discovery,Olech2020period}. Only the HMSFR source G323.46-0.08 is periodic and has experienced an accretion event \citep{proven2019discovery,MacLeod2021g323}.

The HMSFR G188.95+0.89 (also known as S\,252 or AFGL\,5180) is well studied at multiple wavelengths. \citet{oh2010vera}
reported the parallax distance as 1.76$\pm$ 0.11 kpc, Perseus spiral arm \citep{reid2009trigonometric}. The 6.7\,GHz \ch~masers of this source were discovered by \citet{menten1991discovery} and reported periodic, $\tau=$395$\pm$8\,d, by \citet{goedhart2004long,goedhart2014}. 
\citet{kurtz2000hot} report an associated unresolved UCH11 region. \citet{minier2005star} did not detect the continuum radio source, but reported that the masers are projected on a bright mm source with an estimated mass in MM1 of 50$M_{\sun}$. They also detected the presence of CH$_{3}$CN and C$^{18}$O towards the G188.95+0.89 methanol maser site indicating that they are within hot molecular cores (HMCs) with a gas density $\gid$ 10$^{5}$ cm$^{-3}$.

In this paper, we present the results of ongoing 6.7\,GHz \ch\/ maser monitoring for G188.95+0.89. Significant variations of maser features are analysed. Also high-resolution mm-wavelength dust continuum and molecular line emission observations are presented. 

\section{Observations and  Data Reduction}
\subsection{Single-dish Observations}

The radio observations  were made using the 26-m telescope of the Hartebeesthoek Radio Astronomy Observatory
(HartRAO\footnote{See http://www.hartrao.ac.za/spectra/ for further information.}). The 4.5\,cm receiver is comprised of dual, cryogenically cooled, RCP and LCP feeds.  Each polarisation was calibrated independently relative to Hydra~A and 3C123, assuming the flux scale of \citet{ott1994updated}.

All observations employed frequency-switching (FS) and within 2\,hr of zenith. Observations completed with a 1\,MHz bandwidth and recorded with a 1024-channel per polarisation spectrometer (in FS 512 channels are saved).  The velocity resolution achieved is 0.044\kms\/ and the typical 3$\sigma$ root-mean-square (rms) noise per observation was $\sim$1\,Jy. This sensitivity is improved when across several channels, e.g. integration over ten channels improves the sensitivity to $\sim$0.3\,Jy\,\kms. For all the epochs, half-power beam-width pointing correction observations were carried out. The rest frequency was corrected for the Local Standard of Rest (LSR) velocity $v_\textrm{LSR} =$+10.0\kms. Listed in Table \ref{tab:HartObsparam} are the parameters of the telescope, receiver, and observation set-up for each epoch.

Observations were made between 2003 June 30 to 2021 January 18 with a cadence of between 10 to 20\,d; the cadence varied during several flares. No spectroscopic observations were taken between between 2008 September and 2010 December when the 26\,m antenna underwent repairs \citep{Gaylard10}. The position employed at HartRAO is R.A.~=~06$^{h}$~08$^{m}$~53\fs3 and Dec.~=~$+21\degr$~38$\arcmin$~30 (J2000).

\begin{table}
\caption{Summary of single-dish maser monitoring observations at HartRAO.}
\label{tab:HartObsparam}
\begin{tabular}{ll}
\hline
Parameter & Quantity \\
\hline
Receiver & 4.5 cm\\
Maser Transition & J = 5$_{1}$ - 6$_{0}$ A $^{+}$ \\
Rest frequency & 6.668518 GHz\\
System temperature & 57 K\\
Beam width & 7\arcmin\\
Correlator bandwidth & 1.0 MHz \\ Observing mode & Frequency-switching\\
Number of spectral channels & 512\\
Velocity range & 22.5 \kms\\
Correlator resolution & 0.044 \kms\\
Central velocity & +10.0 \kms\\
Sensitivity 3$\sigma$ rms & $\sim$1.0 Jy \\
Monitoring period & 2003 Jun 30 - 2021 Jan 18 \\
Down time for repairs & 2008 Sept - 2010 Dec \\
 \hline
\end{tabular}
\end{table}
\subsection{ALMA Observations}
We obtained ALMA band 6 archival data on G188.95+0.89 (Project ID:2015.1.01454.S) taken on 2016 April 23 (42 antennas of the 12-m main array), 2016 September 17 (38 antennas of the 12-m main array). Observations of G188.95+0.89 (phase tracking center at R.A.$_{(J2000)}$~=~06$^{h}$~08$^{m}$~53\fs3 and Dec.$_{(J2000)}$~=~$+21\degr$~38$\arcmin$~30\arcsec) were carried out at 1.3 mm (230 GHz) with ALMA band 6 in 2015 using two different configurations. The total time on-source was 19 and 30 minutes respectively, and the projected baselines ranged from 15 - 2600 m. The ALMA correlator was configured to cover nine spectral windows (spws). The raw visibility data were calibrated using the Common Astronomy Software Applications (CASA 5.4) standard calibration and imaging tasks. Bandpass and flux calibration were conducted using the sources J0510$-$1800 and J0750$-$1231, respectively. J0603$-$21591 was used as the gain calibrator. The calibrated visibility data from the two observation blocks were combined using CASA task $CONCAT$, and $CLEAN$ was used to produce the images of the continuum (rms of 2.2 mJy\,beam$^{-1}$). Also images of molecular line emission (typical rms of 2.2 mJy\,beam$^{-1}$) with spectral resolution of 1\kms\/ are produced. In this paper, we will focus on the $^{12}$CO (2-1), CH$_3$OH (4$_{(2,2)}$-3$_{(1,2)})$, SiO (5-4) and \c18o (2-1).   
For continuum subtraction, only line-free parts of the spectra were used.  We used CASA to do all the calibration and imaging of the data cubes as well as self calibration to remove residual phase and flux calibration errors. The line data were then imaged with a robust weighting of 0.5.

\section{Results}
\subsection{6.7 GHz \ch~Masers Variability} 

\begin{figure*}
\includegraphics[width=0.99\columnwidth]{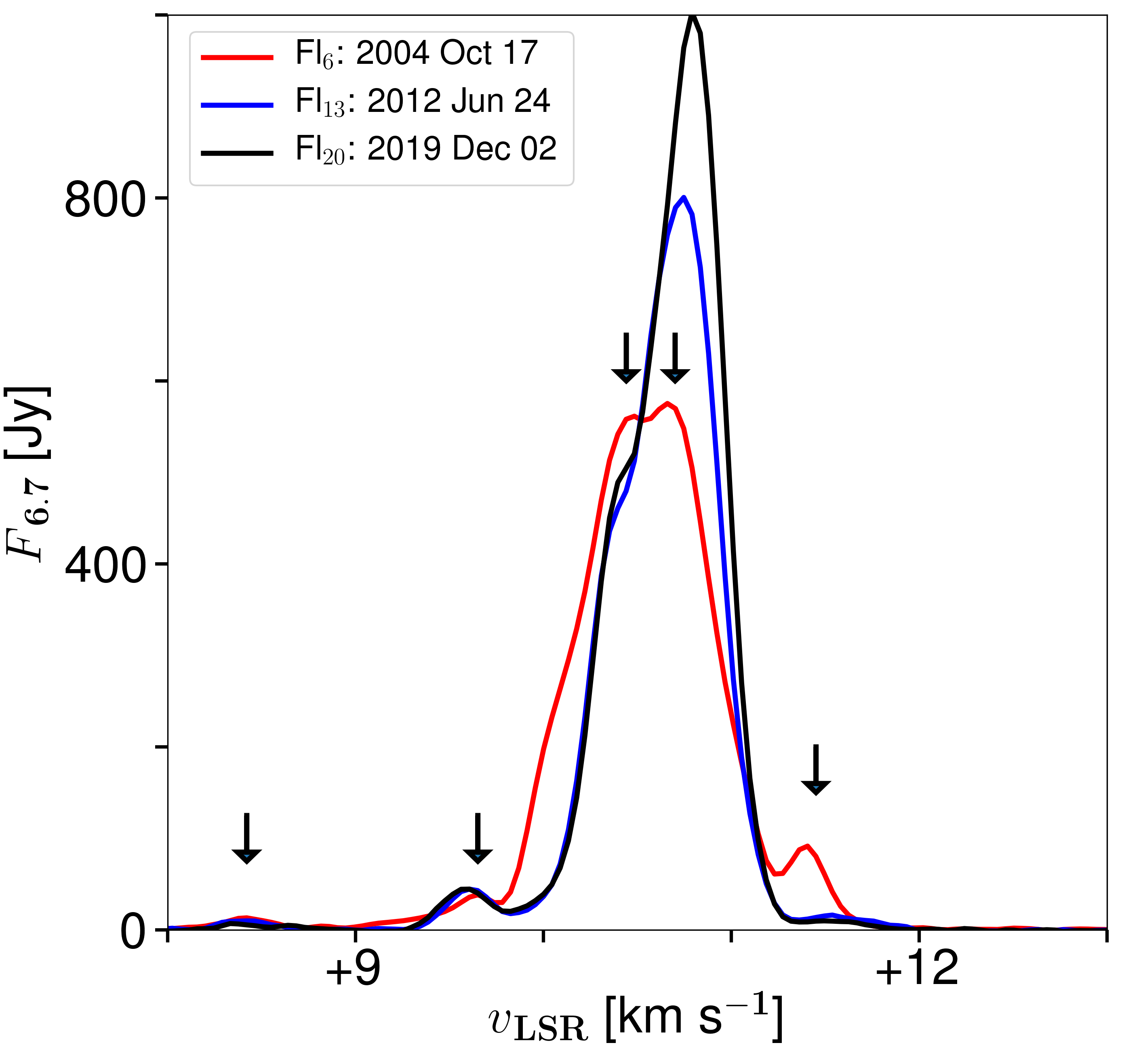}
\caption{Spectra of the 6.7\,GHz methanol masers associated with G188.95+0.89 observed at the maxima of Fl$_{6}$, Fl$_{13}$, and Fl$_{20}$. Arrows mark the selected velocity features studied here.}
\label{fig:methanol_spectra}
\end{figure*}

In Fig. \ref{fig:methanol_spectra} we present selected spectra from the seventeen flare cycles (Fl$_{n}$ for n$=$5 to 21) in the 6.7 GHz methanol maser observations associated with \g188\/ studied here (the original four are published in \citet{goedhart2004long}). The spectra presented are the maxima of three of these flare cycles (flare number Fl$_{n}$ for n$=$6, 13, and 20). In this image it can be seen the evolution of the spectral profile over the 18 years of observations presented. This profile is certainly comprised of many line blended masers, in particular for +10 $< v_{LSR} <$ +11.3\kms. 

\begin{figure*}
\includegraphics[clip,width=\textwidth]{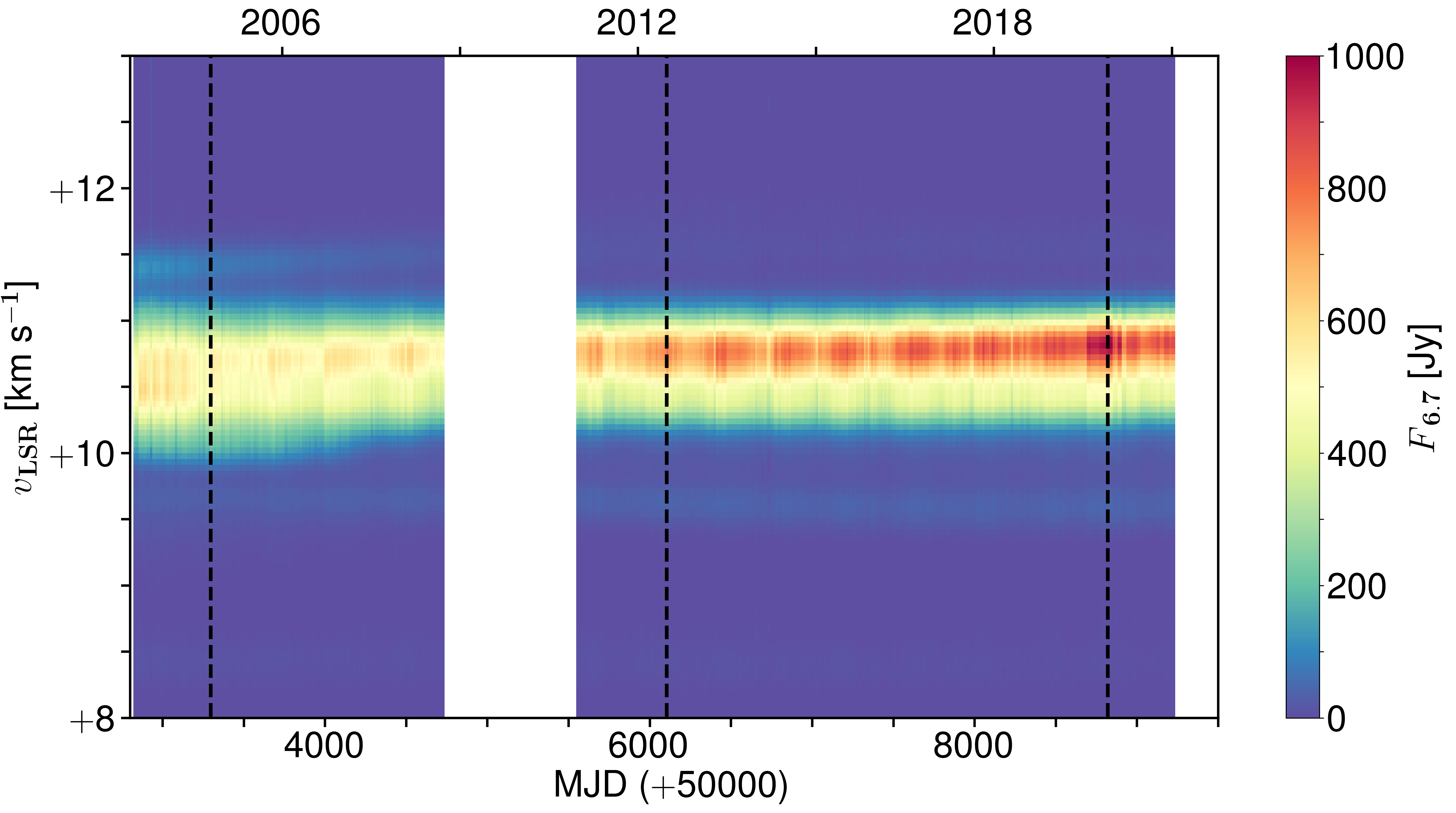}
\caption{The dynamic spectrum of the 6.7\,GHz methanol masers associated with G188.95+0.89. Dashed lines demarcate the spectra of the maximum in Fl$_{6}$, Fl$_{13}$, and Fl$_{20}$ and plotted in Fig. \ref{fig:methanol_spectra}. No observations were taken between 2008 September and 2010 December during repairs.}
\label{fig:dynamic_spectra}
\end{figure*}

A dynamic spectra is the best way to visualise variability of these 6.7 GHz methanol masers associated with \g188, see Fig. \ref{fig:dynamic_spectra}. Features have weakened others strengthened, most appear periodic and variations in velocity, velocity drifts, are apparent. For the purpose of analysis of these observations five features, two at $v_{LSR} =$+10.44 and +10.70\kms\/ in the heavily line merged velocity regime, and others at $v_{LSR} =$+8.42, +9.65, and +11.45\kms\/ are selected. Note, the former two features are likely comprised of multiple masers; the others less likely.

\begin{figure*}
\includegraphics[width=0.99\textwidth]{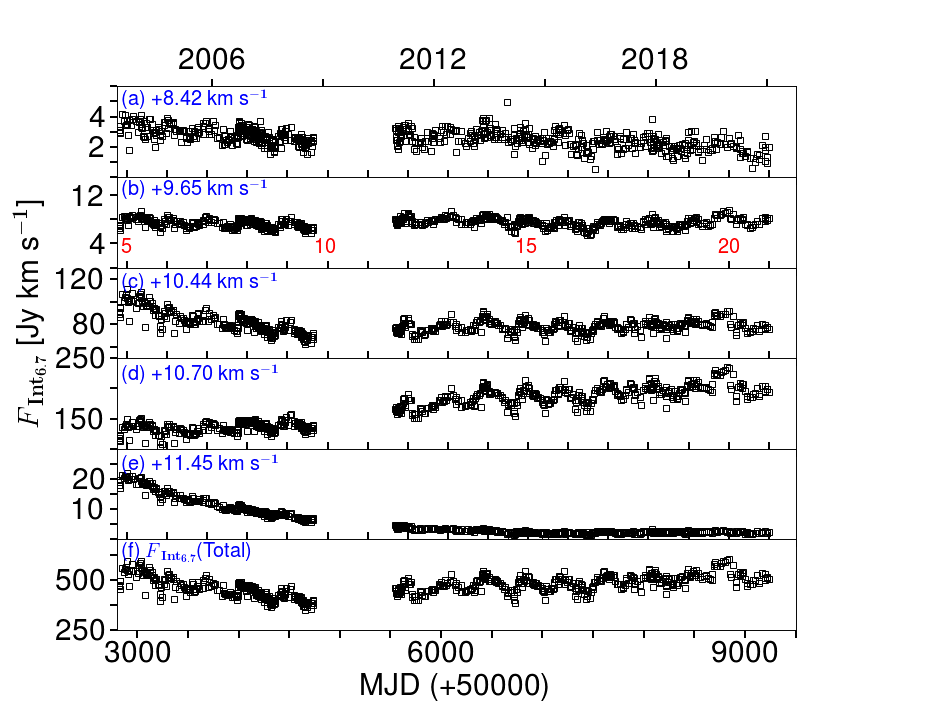}
\caption{Time series plots of $F\mathbf{_{Int_{6.7}}}$ for five maser features identified in Fig. \ref{fig:methanol_spectra} and labelled in each panel. In (d) the total integrated flux density of all masers associated with G188.95$+$0.85 is plotted. Note the x-axis in (b) through (e) denote the estimated maxima of each flare, Fl$_{n}$ for n$=$5 to 21 (for $\tau\sim$395\,d).}
\label{fig:timeseries}
\end{figure*}

Time series plots of the integrated flux density, $F\mathbf{_{Int_{6.7}}}$, for each selected feature are shown in Fig. \ref{fig:timeseries}. A time series plot of the total integrated flux density, $F\mathbf{_{Int_{6.7}}}$(Total), is also shown. For each feature, in each epoch of observation, the velocity associated with the maximum flux density in the stated velocity extent is determined. Linear regression analysis is applied to each velocity time series to determine the velocity drift, and the results are included in this table. Two of the features have measurable velocity drifts, with "goodness of fit" values R$^{2} >$ 50$\%$. The apparent drift seen in the $v_{LSR}=$+10.70\kms\/ feature (see Fig. \ref{fig:methanol_spectra}) may be the result of variations of heavily line merged masers contributing to the feature. The features at $v_{LSR}=$+9.65 and +11.45\kms\/ are best fit by a single Gaussian profile each; both possibly devoid of line merged features.

All five features are periodic though it is not obvious for the feature $v_{LSR} =$+11.45\kms\/ in Fig. \ref{fig:timeseries}. The period of each is determined using, firstly the programme Period04 developed by \citet{lenz2004period04} and secondly the Lomb-Scargle (LS) periodigram \citep{scargle82}. The results are listed in Table \ref{tab:g188_meth}.  The average period for each method is included and are within 3$\sigma$ standard deviation of each other, $\tau_{Period04}=$395$\pm$1\,d against $\tau_{LS}=$397.6$\pm$0.7\,d. 

\begin{figure*}
\includegraphics[width=0.99\textwidth]{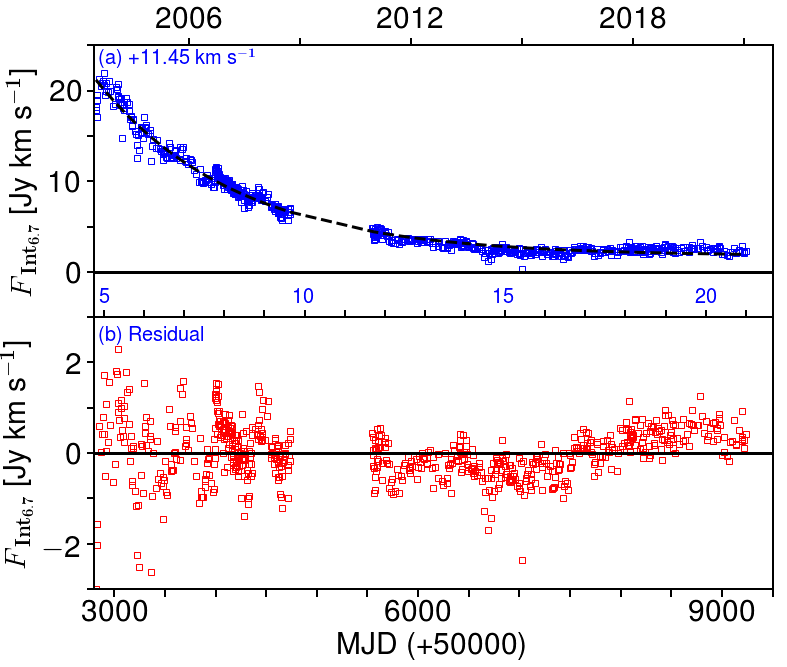}
\caption{ (a) Time series plot of $F\mathbf{_{Int_{6.7}}}$ for the $v_{LSR} =$+11.45\kms\/ feature. The fitted exponentially decaying function (black dashed line) is included. (b) The residual of the integrated flux density, raw data less the fitted exponentially decaying function, is plotted. Note the secondary x-axis in (b) denotes the estimated maxima of each flare, Fl$_{n}$ for n$=$5 to 21 (for $\tau\sim$395\,d).}
\label{fig:fitted}
\end{figure*}

The $v_{LSR}=$+8.42\kms feature is decaying between 2003 and 2021 while the feature $v_{LSR}=$+10.70\kms\/ is increasing in this period. The $v_{LSR}=$+9.65\kms\/ feature appears flat in Fig. \ref{fig:timeseries}, however its associated flux density of the central velocity channel in the velocity range (not plotted) is rising.  When first detected in 1991 \citep{menten1991discovery} the +10.44\kms\/ feature was the brightest feature, here it is second strongest and is decaying until MJD 5000, thereafter it is flat or slightly increasing. The increase is possibly caused by contributions from the +10.70\kms\/ feature.  Finally, for $v_{LSR} =$+11.45\kms\/, it is possible to fit an exponentially decaying function to the flux density time series, see Fig. \ref{fig:fitted}. The fitted function is:

\begin{equation}\label{decay}
   F\mathbf{_{Int_{6.7}}} = a \times e^{MJD/b} + c,
\end{equation}

\noindent where $a=$150$\pm$4\,Jy \kms, $b=-$1380$\pm$20, and $c=$1.74$\pm$0.05\,Jy \kms. \citet{goedhart2004long} reported this feature reached a maximum on 2001 November 12 (MJD 2225), $F_{Int}\sim$45\,Jy \kms. From Eqn. \ref{decay}, on MJD 2225, it is estimated $F_{Int}\sim$30\,Jy \kms. The projected value is significantly lower than the actual value; they are outside estimated errors. The difference in velocity resolution, here it is 0.044\kms while in \citet{goedhart2004long} it is 0.056\kms, may account for this difference. Line-merged features may also be a factor. Some other features varied linearly and did not require de-trending for analysis. The complex varying feature $v_{LSR}=$+10.70\kms\/ will be analysed more elsewhere.

After Eqn. \ref{decay} is subtracted from the original time series the remaining residual, also plotted in Fig. \ref{fig:fitted}, reveals its possible periodic nature. Note the slight increase after MJD 7500; possibly resulting from variations in the $V_{LSR}=$+10.70\kms\/ feature. The total integrated flux density plotted in Fig. \ref{fig:timeseries} is seen first falling prior to MJD 5000 then rising suggesting the features decaying dominate prior to MJD 5000 thereafter brightening of the +10.70\kms\/ feature dominates.

The relative amplitude variation of each flare can be determined using R$_{amp}$:

\begin{equation}\label{flamp}
  R_{amp}=\frac{F_{max}-F_{min}} {S_{min}}=\frac{F_{max}}{F_{min}}-1,
\end{equation}

\noindent where $F_{min}$ and $F_{max}$ are the minimum and maximum flux densities for each flare cycle and each feature. The results of R$_{amp}$ for each velocity feature are included in Table \ref{tab:g188_meth}. This value ranges from 0.1 to 2.8 for the five features. Note that R$_{amp} \sim$1 for the $v_{LSR}=$ +11.45\kms\/ feature; the amplitude is deceasing approximately proportionately during the decay. No phase lags between maser features were found. This may infer that the periodic features are located such that when the cause of variations occur, it affects each nearly simultaneously. There may be lags shorter than our cadence that we cannot measure nor speculate about.

\begin{table*}
\caption{Information of individual 6.7\,GHz maser features. Included are the feature central velocity of each, the velocity extent in which $F_{Int}$ is determined, the velocity drift (1$\sigma$ standard deviation in parenthesis), its goodness of fit (R$^2$), the period determined from the Period04, and LS methods, and description of long-term variation. The average period for each method is presented (1$\sigma$ standard deviation in parenthesis). Comments on velocity drift are given: uncertain (R$^2 <$50$\%$), blue-shifting, and red-shifting. Comments on general trends of the flux density of each feature are given (each are periodic).}
\label{tab:g188_meth}

\begingroup
\setlength{\tabcolsep}{6pt} 

\begin{tabular}{ccccccccc}
\hline

\multicolumn{4}{c}{Velocity} &
\multicolumn{2}{c}{Period} & Relative& \multicolumn{2}{c}{Comments} \\

Feature & Extent & Drift & R$^{2}$ &   $\tau_\textrm{Period04}$ & $\tau_\textrm{LS}$ & Amplitude & Velocity & Flux density\\

(\kms) & (\kms) &  ($\times$ 10$^{-5}$ \kms\,d$^{-1}$)& ($\%$) & (d) & (d) & & &\\
\hline
$+$8.42 & 0.48   &$-$1.18(7)  &22.9 & 394.4 &397.0 & 2.8 & uncertain&slowly decreasing \\

$+$9.65 & 0.26   & $-$2.38(3)& 81.6& 395.8  &396.7 & 0.6 &blue-shifting & slowly increasing \\

$+$10.44 & 0.18  & +0.42(4)& 30.4 & 393.5 &398.3 & 0.1 &uncertain &complex\\

$+$10.70 & 0.31   & $+$0.39(5) &5.7 & 396.3 &398.2 & 1.0& uncertain &increasing \\

$+$11.45 & 0.31   & $+$1.88(4)&66.0 &  & 397.8 & $\sim$1 & red-shifting& exponential decay \\
\hline
Average & & & & 395(1) & 397.6(7) &  &\\
\hline
\end{tabular} 
\endgroup
\end{table*}

\subsection{\g188\/ Continuum Emission}

\begin{table*}
	\centering
	\caption{Parameters of the detected dust cores}
	\label{tab:cont parameters}
	\begin{tabular}{lcccccr} 
		\hline
		Object-name & R.A. & Dec. & Peak flux & Integrated flux &  V$_{sys}$ & Core mass\\ 
            & (h m s) & ($ ^{\circ}$~$\prime$~$\prime\prime$) & (mJy~beam$^{-1}$) & (mJy) & (km s$^{-1}$) & (M$_{\odot}$)\\ 
		\hline
MM1 & 06 08 53.33 & 21\ 38\ 28.9 & 71.7 & 71.1 & 5.0 & 8.2 \\
MM2 & 06 08 53.49 & 21\ 38\ 30.5 & 22.5 & 41.3 & 2.0 & 4.8 \\ 
MM3 & 06 08 54.13 & 21\ 38\ 34.4 & 5.6 & 9.4 & 3.0 & 1.1 \\
MM4 & 06 08 52.86 & 21\ 38\ 29.5 & 3.2 & 3.8 & 4.5 & 0.5 \\
MM5$^{*}$ & 06 08 53.35 & 21\ 38\ 11.6 & 8.3 & 14.2 & 2.0 & 1.4 \\
MM6$^{*}$ & 06 08 53.42 & 21\ 38\ 13.7 & 1.9 & 3.6 & 4.0 & 0.4 \\
MM7$^{*}$ & 06 08 53.23 & 21\ 38\ 09.7 & 9.8 & 16.7 & - & 1.6 \\ 
MM8$^{*}$ & 06 08 52.97 & 21\ 38\ 11.1 & 5.8 & 6.0 & 4.0 & 0.6 \\ 
		\hline
	\end{tabular}
\end{table*}

\begin{figure*}
        \centering
        \begin{subfigure}[b]{0.95\textwidth}
            \centering
            \includegraphics[width=0.9\textwidth]{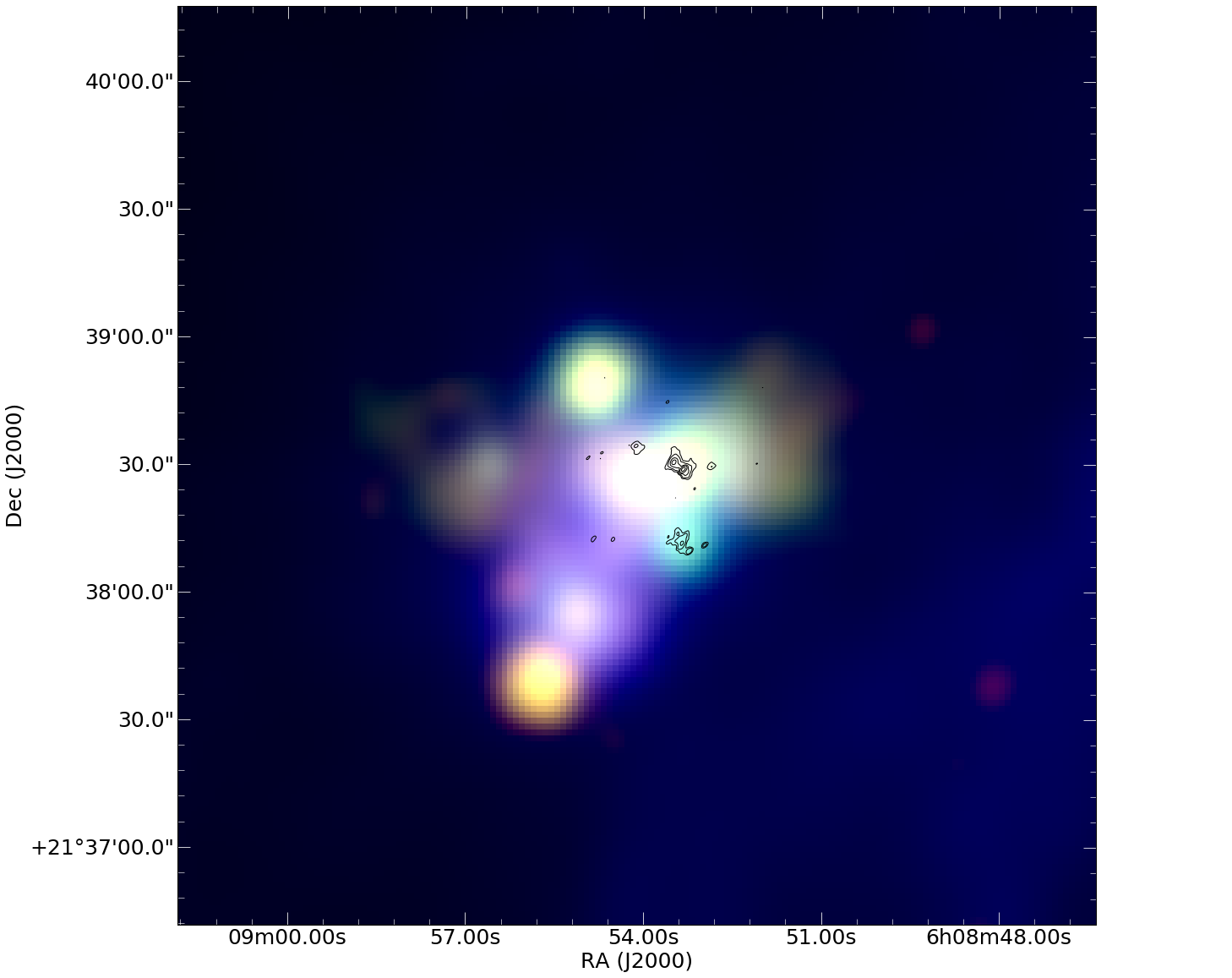}
        \end{subfigure}
        \hfill
        \begin{subfigure}[b]{1\textwidth}
            \centering
            \includegraphics[width=1.0\textwidth]{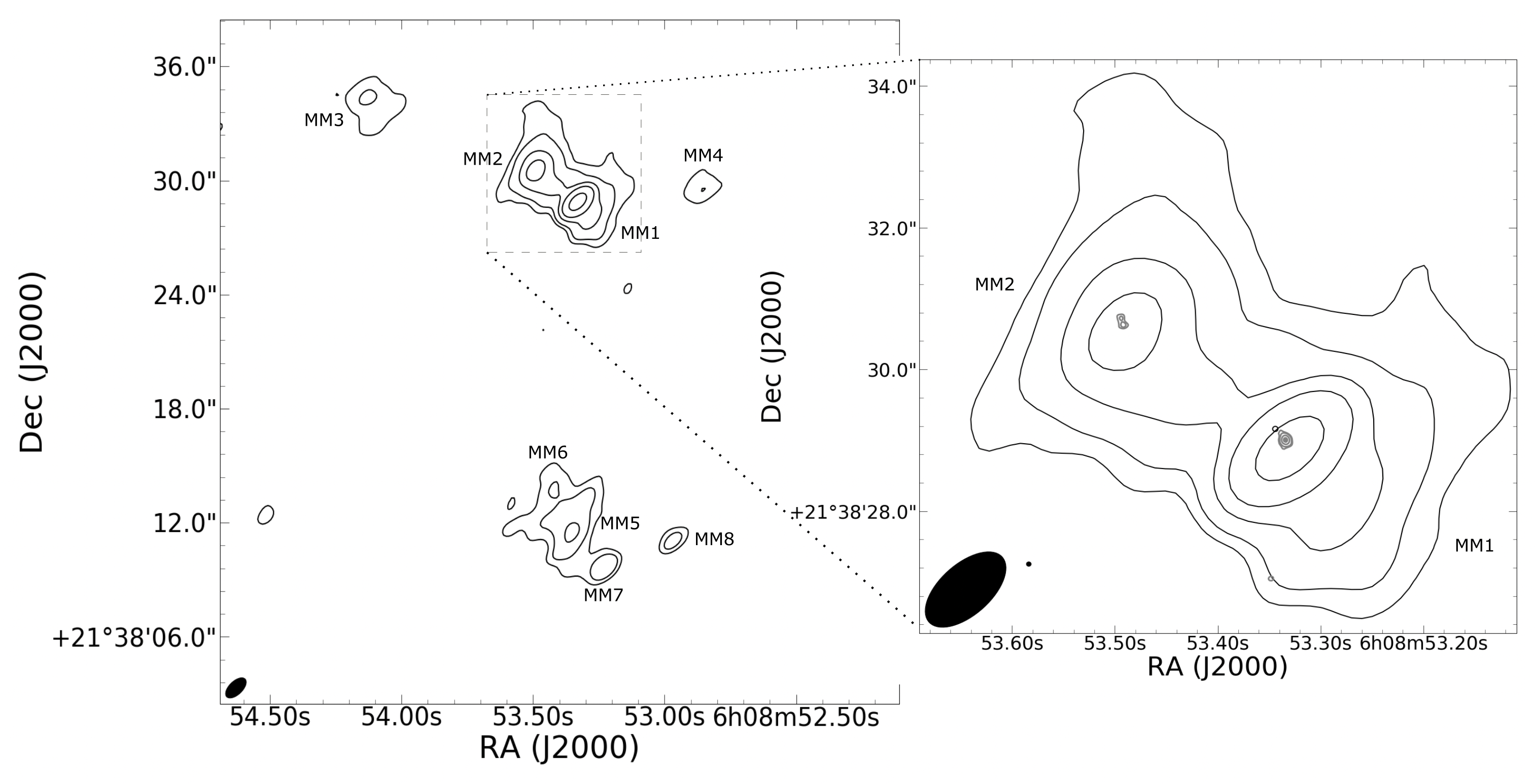}
        \end{subfigure}
        \hfill

        \caption{ \small(top) Composite 3-color WISE image of \g188. WISE band 1 (3.4\,$\mu$m), 2, (4.6\,$\mu$m) and 3 (12\,$\mu$m) are represented red, green and blue. (bottom) ALMA band 6 (black contour, levels = [2, 4, 10, 20, 50] mJy\,beam$^{-1}$) and in the zoom-in, band 7 (gray contour, levels = [2, 4, 20, 60, 80] mJy\,beam$^{-1}$) dust continuum emission of \g188. The filled ellipses on the bottom left corner beams of the band 6 (larger ellipse) and band 7 (smaller ellipse). The open circle near MM1 peak indicates the position of the periodic 6.7 GHz \ch~masers.}
        \label{composite}
    \end{figure*}

In Fig. \ref{composite} (top panel), we show the composite image of \g188~in WISE bands 1\,(3.4$\mu$m: red), 2 (4.6\,$\mu$m: green) and 3 (12\,$\mu$m: blue) and ALMA 1.3\,mm dust continuum emission. The WISE image show a central infrared source (bright in all 3 bands) corresponding to the 850\,$\mu$m SCUBA MM1 object, and a green (4.6\,$\mu$m) dominant object south of the central object corresponding to the 850\,$\mu$m SCUBA MM2 object \citep{minier2005star}. \g188~is resolved into eight 1.3\,mm objects (MM1--MM8) with ALMA. MM1--MM4 and MM5--MM8 objects (Fig. \ref{composite}, bottom panel) are associated with 850\,$\mu$m SCUBA MM1 and MM2 objects of \citet{minier2005star}, respectively.

ALMA 1.3\,mm MM1 and MM2 are central objects (Fig. \ref{composite} bottom zoom-in) and MM1 (the brightest dust continuum object) is associated with the periodic 6.7\,GHz \ch~maser source. Interestingly, a recent ALMA band 7 observations resolved the continuum core in MM1 into a single object but resolved MM2 into 2 continuum cores (gray contours of Fig. \ref{composite} bottom zoom-in). Details of the detected continuum sources and their masses are presented in Table \ref{tab:cont parameters}.

The dust mass  $M_{d}$ in Table \ref{tab:cont parameters}, can be estimated by assuming optically thin dust emission using \citet{Hildebrand1983}:
\begin{equation}
M_{d} = \dfrac{S_{\nu} \, D^{2}}{\kappa_{\nu} \, B_{\nu}(T_{d})} \label{mass equation} 
\end{equation}
The dust continuum flux density $S_{\nu}$ is given in Table \ref{tab:cont parameters} at frequency, $\nu$ with $D$ the distance to the source which is 1.76 kpc and $\kappa_{\nu}$ the dust opacity
per unit mass, $\kappa_{\nu}$ $ =$ 0.33 cm$^{2}$ g$^{-1}$ for 230 GHz \citep{weingartner2001dust}. $B_{\nu}(T_{d}) $ is the Planck function at dust temperature, $T_{d}$. The temperature used in the dust mass estimations were obtained from \citet{minier2005star}, for cores MM1 to MM4 a temperature of 42 K is used and for cores MM5 to MM8, 50 K is used. To obtain the mass of the cores a  a gas-to-dust mass ratio 100 was used and the results are given in Table \ref{tab:cont parameters}

The integrated flux densities for MM5 to MM8, that were also used in the core mass estimates, are not primary beam corrected since MM5--MM8 are located close to the edge of the primary beam and will not be discussed in detail due to limited sensitivity.

\begin{table}
\setlength{\tabcolsep}{1pt} 
\renewcommand{\arraystretch}{1} 
 \caption{Note. Columns are species, transition, rest frequency, energy of the upper level and number of atoms present.}
 \label{tab:5}
 \begin{tabular}[h!]{lcccc}
  \hline
  Observed molecular species\\
  \hline
  Molecule & Transition & Rest frequency & E$_{L}$ & E$_{U}$\\
  &  & (GHz) & (K) & (K)\\
  \hline
  SiO        & 2-1             & 217.10498000 & 20.84 & 31.26 \\
  C$^{18}$O  & 2-1             & 219.56035410 & 5.27 & 15.81\\
  CH$_{3}$OH & 4$_{(2,2)}$-3$_{(1,2)}$ & 218.44006300 & 34.50 & 45.46 \\
  $^{12}$CO  & 2-1             & 230.53800000 & 5.53 & 16.60 \\
  \hline
 \end{tabular}
\end{table}

\subsection{\g188\/ Millimeter Line Emission}
A number of thermal molecular lines were detected towards the 8 millimeter continuum objects, however, in this paper we will focus only the \ch~(4$_{(2,2)}-$3$_{(1,2)}$), SiO ($J=2-1$), $^{12}$CO ($J=2-1$) and C$^{18}$O ($J=2-1$).\\

\begin{figure*}
            \includegraphics[width=1.3\textwidth]{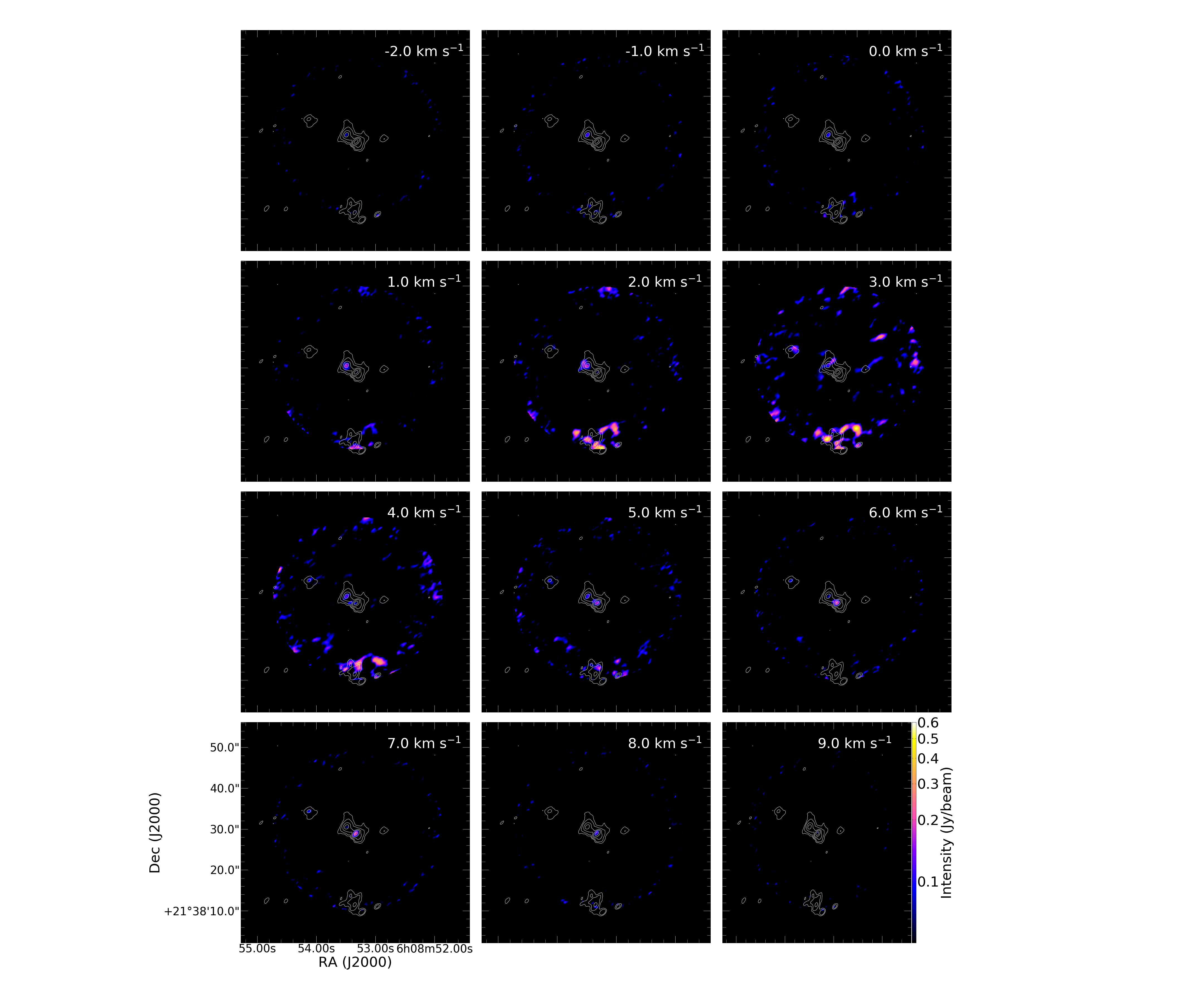}
 \caption{\ch~4(2,2)-3(1,2) thermal line channel map with dust continuum overlaid (black contour, levels = [0.002, 0.004, 0.02, 0.06, 0.08] Jy\,beam$^{-1}$) dust continuum emission of \g188. The color scale is the intensity in Jy\,beam$^{-1}$.}
        \label{ch3oh}
    \end{figure*}

\noindent {\bf \ch~4$_{(2,2)}$-3$_{(1,2)}$}\\
Emission from the \ch~(4$_{(2,2)}-$3$_{(1,2)}$) line was detected towards all 8 cores, see Fig. \ref{ch3oh}. The emission is strongest towards the southern MM5--MM8 and weakest towards MM4. The nature of the emission towards MM5--MM8 is contaminated by the effect of the decreased sensitivity towards the edge of the primary beam. Beyond detection or non-detection of an emission, no further discussions will be made for these objects.

One interesting feature of the \ch~(4$_{(2,2)}$-3$_{(1,2)})$ emission towards MM2 is the presence of a double peak in the emission at the systemic velocity of MM2. There is \ch~(4$_{(2,2)}-$3$_{(1,2)}$) emission associated with MM1. \citet{minier2005star} indicated the temperature of the dust of the clump hosting the 1.3\,mm ALMA MM1--MM4 sources to be $\sim$150\,K.\\

\begin{figure*}
          \includegraphics[width=1.3\textwidth]{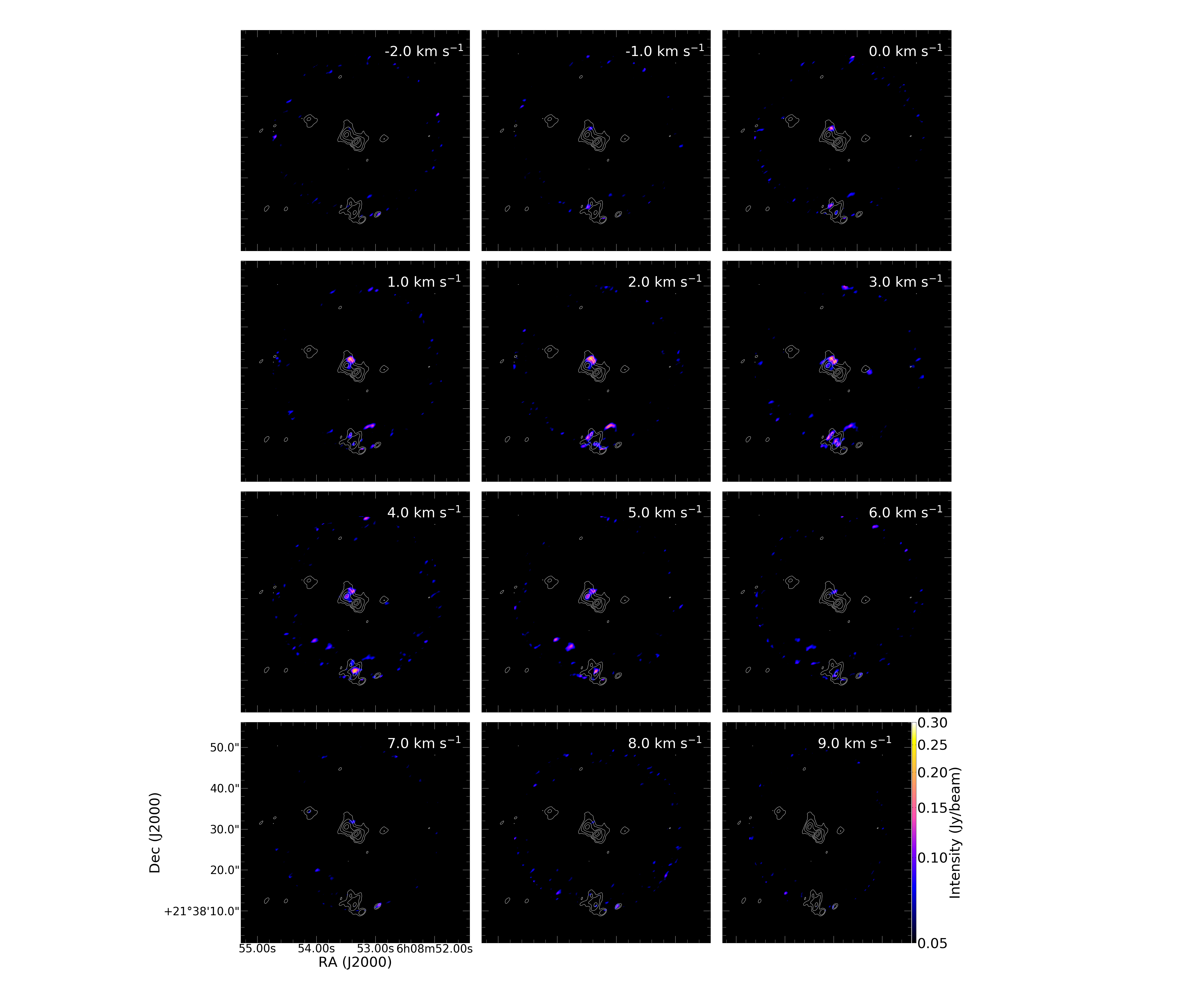}
 \caption{SiO~($J=2-1$) thermal line channel map with dust continuum overlaid (black contour, levels = [0.002, 0.004, 0.02, 0.06, 0.08] Jy\,beam$^{-1}$) dust continuum emission of \g188. The color scale is the intensity in Jy\,beam$^{-1}$.}
        \label{sio}
    \end{figure*}

\noindent {\bf SiO ($J=2-1$)}\\
Emission of SiO ($J=2-1$) is known to trace shocks especially in star-forming regions. The dominant SiO emission is observed in the north-west region of MM2, see Fig. \ref{sio}. The emission has a bow-shock morphology in the channel corresponding to the systemic velocity of MM2. There are SiO emission (in some cases, very faint emission) associated with each of MM3 to MM8, however, no SiO emission is detected towards the MM1 object.\\

\begin{figure*}
    \includegraphics[width=1.3\textwidth]{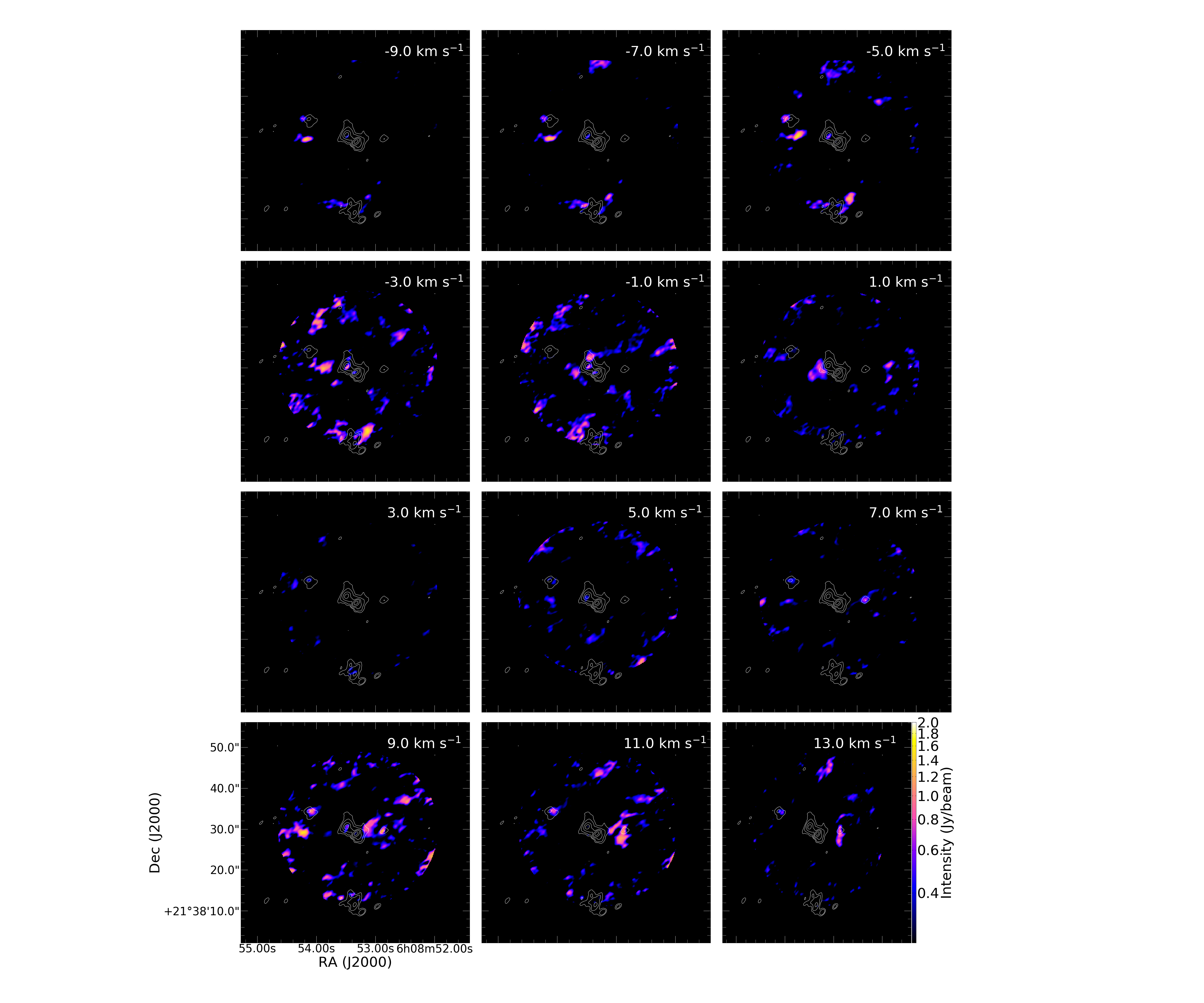}
 \caption{$^{12}$CO ($J=2-1$) channel map with dust continuum overlaid (black contour, levels = [0.002, 0.004, 0.02, 0.06, 0.08] Jy\,beam$^{-1}$) dust continuum emission of \g188. The color scale is the intensity in Jy\,beam$^{-1}$.}
        \label{c12o}
    \end{figure*}

\noindent {\bf $^{12}$CO ($J=2-1$)}\\
The $^{12}$CO ($J=2-1$) emission in \g188\/ is complicated owing to the effect of the strongly self-absorbed features seen in the line emission. The other plausible reason for the observed complex distribution is that all the millimeter objects may be driving outflows. If the strong emission east of MM2 observed in the $-$3\kms\/ channel and the emission west of MM2 in the 9\kms\/ channel are both associated with MM2, then they may demarcate an east-west bipolar outflow in MM2, see Fig. \ref{c12o}. There is also some emission north of MM2 and could signify a second outflow emanating from the object; suggestive of multiple YSOs in MM2. The other mm objects are all associated with $^{12}$CO ($J=2-1$) emission.\\

\begin{figure*}
          \includegraphics[width=1.3\textwidth]{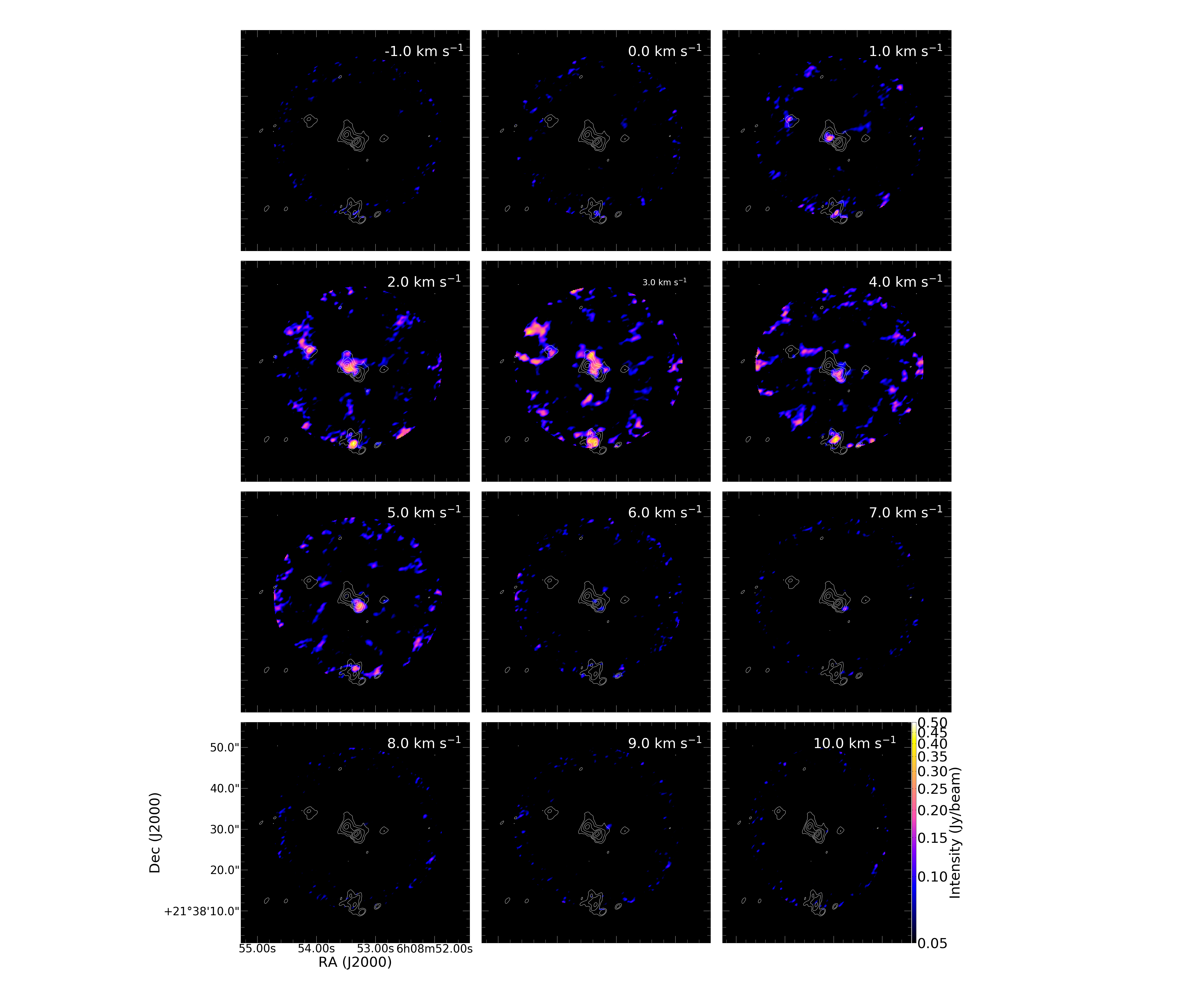}
        \caption{C$^{18}$O ($J=2-1$) channel map with dust continuum overlaid (black contour, levels = [0.002, 0.004, 0.02, 0.06, 0.08] Jy\,beam$^{-1}$) dust continuum emission of \g188. The color scale is the intensity in Jy\,beam$^{-1}$.}
        \label{c18o}
    \end{figure*}

\noindent {\bf C$^{18}$O ($J=2-1$)}\\
We detected high density tracer, C$^{18}$O ($J=2-1$), towards all MM1--MM8, see Fig. \ref{c18o}. MM1 and MM2 are the dominant sources of the C$^{18}$O ($J=2-1$) emission. Interestingly, the brightest C$^{18}$O ($J=2-1$) channel (3\,\kms~channel) show a distribution of the C$^{18}$O ($J=2-1$) emission to lie in the interface between MM1 and MM2 cores. The possible implication of this to the observed variability in the 6.7\,GHz \ch~masers will be discussed in Section \ref{sec:var}

\section{Discussion}

\subsection{Implications of the variability in the 6.7\,GHz \ch~Masers}
\label{sec:var}

Originally, \citet{goedhart2004long} reported the period of G188.89+0.89 was 416\,d; this was revised in \citet{goedhart2014} to 395\,d. Here, using the longer time series and from two separate methods, the shorter period is confirmed ($\tau_{Period04}=395\pm1$\,d and $\tau_{LS}=397.6\pm0.7$\,d). \citet{Durjasz2021} also confirm, though observing fewer flare cycles, the shorter period, $\tau=395\pm8$\,d.

Class II methanol masers are only found associated with high mass star-forming regions \citep{minier2003masslmt}. Such regions experience accretion events (identified by flaring masers) such as those reported in S255IR-NIRS3 \citep{Fujisawa2015S255,Szymczak2018S255}, NGC6334I \citep{macleod2018masing}, G323.46-0.08 \citep{proven2019discovery, MacLeod2021g323}, and G358.93-0.03 \citep{Sugiyama2019g358, MacLeod2019g358}.  However, three of the features in Table \ref{tab:g188_meth} are decaying, in particular the $v_{LSR}=$+11.45\kms decays exponentially (Fig. \ref{fig:fitted}). Interesting, historical observations suggest the originally detected peak flux density feature, $v_{LSR}\sim$+10.5\kms, has varied markedly since detection, less than 500\,Jy before 1993  \citep{menten1991discovery,caswell1995galactic}, greater than 600\,Jy before 2009 \citep{goedhart2004long, Green2012MXsurvey} to $\sim$500\,Jy on average here. Note at present the peak flux density is found at $v_{LSR}=$+10.70\kms\/ and not +10.44\kms. \citet{caswell1995galactic} reported that the feature $v_{LSR}=$+11.2\kms\/ decreased slightly from 1991 ($\sim$250 \citet{menten1991discovery}) to 1993 ($\sim$230\,Jy). In 1999 August it was only $\sim$100\,Jy \citep{Szymczak2000m67} while reaching a maximum, $\sim$160\,Jy, in 2001 November \citep{goedhart2004long}; thereafter the exponential decay is reported here. Above Eqn. \ref{decay} cannot describe these variations, nor does it fit the maximum reported in \citet{goedhart2004long}. It appears a weak flaring event, increasing by a factor of $\sim$2.6 in flux density, occurred between 1999 August and 2001 November in the periodic feature at $v_{LSR}=$+11.45\kms. This decaying periodic feature resembles that shown for decaying periodic maser features in G323.46$-$0.08 \citep{MacLeod2021g323}. Further analysis of historical data, and data other transitions (.e.g. 12.2\,GHz \ch\/ and/or 22.2\,GHz H$_2$O), are required to confirm.

Spot maps of the 6.7\,GHz methanol masers are shown in \citet{minier2000vlbi, Hu2016spotmap}; all features reside in a 180$\times$180\,au box assuming a distance of 1.76\,kpc \citep{oh2010vera} about the bright reference feature ($v_{LSR}=$+10.33\kms). The weakening features reside within $\sim$50\,au radius of the reference feature, $v_{LSR}=$+8.42\kms\/ slightly north while +11.45\kms\/ is slightly south of the reference maser. The strengthening feature (+10.70\kms) is found $\sim$90\,au to the South-East from the reference maser. The non-varying feature, $v_{LSR}=$+9.65\kms, is further South-East ($\sim$140\,au). \citet{minier2000vlbi} suggest that the masers surrounding the reference maser could be part of an outflow from the disk.

It is not clear whether velocity drifts, including those seen here, are caused by variability of spectrally blended masers \citep{Szymczak2014} or by motion of the gas \citep{goddi2011infall}. The accretion disks surrounding massive star-forming regions may experience infalling gas. \citet{Szymczak2014} report a velocity drifting feature in Cepheus A; they suggest it may be an artifact of variable line-merged features or the result of masers located in regions of infalling gas. \citet{MacLeod2021g962} propose the inner radius of the accretion disk surrounding the protostar of G9.62+0.20E is infalling resulting in systematic velocity drifts of its associated methanol masers. However, they also state it may be caused by masers in a precessing disk. Above in Table \ref{tab:g188_meth} it can be seen two features experience measurable velocity drift. The blue-shifted feature, $v_{LSR}=$+9.65\kms\/ is blue-shifting  ($-2.38\pm0.03 \times 10^{-5}$\kms\,d$^{-1}$) while the other, $v_{LSR}=$+11.45\kms\/, is red-shifting ($+1.88\pm0.04 \times 10^{-5}$\kms\,d$^{-1}$). The drifts are of the order of those seen in G9.62+0.20E \citep{MacLeod2021g962}, ranging from $-$6 to $+$3$ \times 10^{-5}$\kms\,d$^{-1}$. However, an insufficient number of features with measurable velocity drift are found thus rendering any comparison to G9.62+0.20E indeterminate. These two features are found well-separated in the spot maps shown in both \citet{minier2000vlbi, Hu2016spotmap}, by $\sim$200\,au. Perhaps the central features are influenced by an outflow \citep{minier2000vlbi} and this motion may force the features at the edges to experience velocity drift. Even the now brightest feature, at $v_{LSR}=$+10.70\kms, is blue-shifting though this may be due only to variations of heavily line-merged features. All of these periodic masers are associated with MM1. More observation and analysis is required to explain these variations.

\subsection{Binary System in \g188-MM2}
\begin{figure*}
\includegraphics[width=0.9\textwidth]{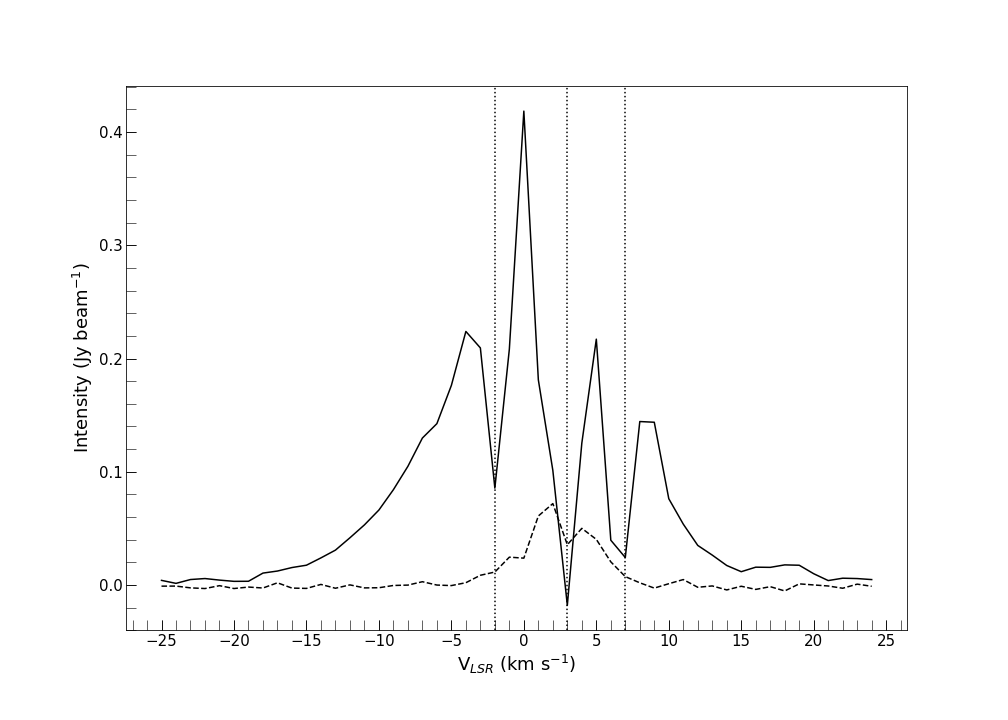}

        \caption{$^{12}$CO (solid lines) and \ch~(dashed lines) spectra (of \g188 extracted with an ellipse of size $\sim$4$"$ centred on MM2. The vertical lines indicate the observed absorption features in $^{12}$CO line. The multiple absorption features in the $^{12}$CO emission and the double peak feature in the \ch~line point to multiplicity (or at least binary) of driving sources in MM2. Band 7 ALMA dust continuum detected 2 cores in MM2.}
        \label{c12o_spec}
    \end{figure*}

There are a number of indicators suggesting the existence of a binary system in MM2. First, the high-resolution ALMA band 7 dust continuum image resolves MM2 into 2 cores separated by 0.1$"$. Second, the \ch~(4$_{(2,2)}-$3$_{(1,2)}$) emission is coincident with the systemic velocity and the spectral profile of emission towards MM2 has two pronounced features (see 3\kms\/ channel of Fig. \ref{ch3oh}). In Figure \ref{c12o_spec}, the spectra of $^{12}$CO (solid lines) and  \ch~(4$_{(2,2)}-$3$_{(1,2)}$) (dashed line) taken from MM2 using a 4$"$ ellipse are presented. It is important to note the $^{12}$CO emission is optically thick. In the $^{12}$CO spectra (at least) self-absorption features are visible, implying the $^{12}$CO emission is absorbed by cold foreground material in the region. The three vertical dotted line in Figure \ref{c12o_spec} indicate the prominent absorption feature (middle line at 3\kms) and two absorption features ($-$2 and 7 \kms) one on each side of the prominent feature. We noted that missing flux issue can give rise to such spectral feature but in this case the observed absorption feature is real.. The most prominent absorption feature in the $^{12}$CO spectra corresponds in \vlsr\/ with the \ch\/ (4$_{(2,2)}-$3$_{(1,2)}$) absorption feature separating the 2 resolved cores in MM2. All these observed features suggest multiple (or at least binary) young stellar objects in MM2.

Our suggestion of the presence of at least a binary, or multiple, YSOs in MM2 is supported by the complexity of the orientation of the outflowing gas traced by $^{12}$CO, see Fig. \ref{c12o}. An east-west outflow feature with an axis through the peak position of MM2, as well as a north-west and south-east outflow also intersecting with MM2, can be observed. These support the suggestion at least a binary system is present in MM2. High-resolution continuum and line observations will be required test the multiplicity of YSOs in MM2.

\section{Conclusion}
We presented results of 1.5 decades of monitoring observations of 6.7\,GHz \ch~masers towards \g188 and ALMA band 6 observational results and found the following;
\begin{enumerate}
\item \g188~protocluster is resolved into eight 1.3 mm objects (MM1-MM8). MM1 and MM2 are the central objects and MM1 (the brightest dust continuum object) is associated with the periodic 6.7 GHz CH$_3$OH  maser source. 
\item MM2 hosts more than one YSO, likely a binary system. 
\item  Strong emission of $^{12}$CO at the east and west of MM2  point to the presence of an east-west bipolar outflow in MM2. Emissions north of MM2 also suggest a second outflow emanating from the object, which could signal multiplicity of YSOs in MM2.
\item All five features are periodic, suggesting a common background source of seed photon. The light curve shape of the maser features are similar to those of Mira variable stars and could suggest pulsation of the protostar as the possible driver of the periodicity \citep{Inayoshi_2013}.
\item  Two \ch~maser features are reported with measurable velocity drift.
\item Outflows are identified in \co\,(2-1) line emission. While no direct detection of accretion disks was possible with the current observations, the detected outflows suggest the presence of accretion disks in  the source.  Accretion disk may experience infalling gas  and velocity drifts  may be due to an artifact of variable line-merged features or by infalling gas.

\end{enumerate}

\section*{Acknowledgements}
MMM acknowledge the support of The Technical University of Kenya, NRF-South Africa, HartRAO-South Africa and the Newton Fund. This paper makes use of the following ALMA data: ADS/JAO.ALMA 2015.101454.S. ALMA is a partnership of ESO (representing its member states), NSF (USA) and NINS (Japan), together with NRC (Canada) , MOST and ASIAA (Taiwan), and KASI (Republic of Korea), in cooperation with the Republic of Chile. The Joint ALMA Observatory is operated by ESO, AUI/NRAO and NAOJ. This publication makes use of data products from the Wide-field Infrared Survey Explorer, which is a joint project of the University of California, Los Angeles, and the Jet Propulsion Laboratory/California Institute of Technology, funded by the National Aeronautics and Space Administration.

\section*{data availability}
\begin{itemize}
\item This paper makes use of the following ALMA data: ADS/JAO.ALMA 2015.101454.S. and can be accessed on the ALMA Science portal.
\item This publication makes use of data products HartRAO 26m maser monitoring project and will be made available on a reasonable request.
 
 \item The WISE data used in this publication can be access on the IRSA public data archive.

\end{itemize}

\bibliographystyle{mnras}
\bibliography{G188} 







\bsp	
\label{lastpage}
\end{document}